\documentclass[12pt]{iopart}
\usepackage[dvipdfmx]{color}
\usepackage[dvipdfmx]{graphicx}
\usepackage{amssymb}
\usepackage{here}

\newcommand{\remove}[1]{}
\newcommand{\com}[1]{}
\newcommand{\blue}[1]{#1}

\begin{document}

%\title{Synchronization and Transition in the Kuramoto Model with Asymmetric Bimodal Distribution of Natural Frequencies} 
\title{Nonstandard transitions in the Kuramoto model: A role of asymmetry in natural frequency distributions}

\author{Yu Terada, Keigo Ito, Toshio Aoyagi and Yoshiyuki Y. Yamaguchi}

\address{Graduate School of Informatics, Kyoto University, Kyoto 606-8501, Japan}
\ead{y-terada@acs.i.kyoto-u.ac.jp, k.ito@acs.i.kyoto-u.ac.jp, aoyagi@acs.i.kyoto-u.ac.jp and yyama@amp.i.kyoto-u.ac.jp}
\vspace{10pt}
\begin{indented}
\item[]\today
\end{indented}

\begin{abstract}
We study transitions in the Kuramoto model by shedding light on asymmetry in the natural frequency distribution, which has been assumed to be symmetric in many previous studies.
The asymmetry brings two nonstandard bifurcation diagrams, with the aid of bimodality.
The first diagram consists of stationary states, and has the standard continuous synchronization transition and a subsequent discontinuous transition as the coupling strength increases.
Such a bifurcation diagram has been also reported in a variant model, which breaks the odd symmetry of the coupling function by introducing the phase lag.
The second diagram includes the oscillatory state emerging from the partially synchronized state and followed by a discontinuous transition.
This diagram is firstly revealed in this study.
The two bifurcation diagrams are obtained by employing the Ott-Antonsen ansatz, and are verified by direct $N$-body simulations.
We conclude that the asymmetry in distribution, with the bimodality, plays a similar role to the phase lag, and diversifies the transitions.
\end{abstract}

%!TEX encoding = UTF-8 Unicode
%\maketitle

\section{Introduction}

Rhythmic phenomena are ubiquitous in various fields, such as biology, chemistry, engineering, physics and so on \cite{kuramoto1984chemical,strogatz2003sync,winfree2001geometry,pikovsky2003synchronization}.
Rhythmic phenomena are classified into the individual and collective levels, and individual rhythmic units cooperate to synchronize and organize the collective rhythmic motion.
The individual rhythmic unit is mathematically described by a dynamical system having a limit cycle, and a limit-cycle oscillator is reduced to a phase oscillator by using the phase reduction technique \cite{kuramoto1984chemical}, which extracts the rhythmic part from the limit-cycle oscillator.
The phase reduction technique is applicable even under interactions with other oscillators as perturbations, and we can consider coupled phase-oscillators, which describes the synchronization.

The Kuramoto model is a paradigmatic phase-oscillator system, which has the all-to-all, mean-field interaction \cite{kuramoto1975self,strogatz2000kuramoto,acebron2005kuramoto,gupta2014kuramoto}.
In this model, each oscillator has the so-called natural frequency, which is constant in time and obeys a probability distribution.
The coupling function between a pair of oscillators is represented by the sine function, where the argument is just the phase difference between the coupled oscillators.
This simple model describes the transition from the incoherent state to the partially synchronized state by strengthening the couplings even when the natural frequencies are not identical.
This transition is called the synchronization transition, and is continuous when natural frequency distributions are symmetric and unimodal as originally considered \cite{kuramoto1975self}.
The steady states (the incoherent and the partially synchronized states for instance) and the transitions are represented on bifurcation diagrams depending on parameters like the coupling strength.
The bifurcation diagrams depend on the natural frequency distributions as well as the coupling function, and drawing bifurcation diagrams is one of the central issues in the coupled phase-oscillator systems as the phase diagrams in condensed matters.

The bifurcation diagram crucially depends on symmetry of the system as illustrated by the simple dynamical system $\dot{x}=x-x^3-h$ \cite{strogatz2014nonlinear,golubitsky1985singularities}.
In the case $h=0$, where the equation is invariant with respect to the transformation $x\to-x$, the system undergoes the pitchfork bifurcation.
On the other hand, when $h\neq0$, the equation is asymmetric and the system exhibits not the pitchfork but the saddle-node bifurcation.
Similarly, the Kuramoto model shows different bifurcation diagrams from the originally obtained one by breaking symmetry of the coupling function or of the natural frequency distributions.

We first turn our attention to the coupling function.
In the Sakaguchi-Kuramoto model \cite{sakaguchi1986soluble}, which introduces the phase lag parameter in the sine coupling function of the Kuramoto model, the non-zero lag breaks the odd symmetry of the coupling function and the discontinuous transition can appear \cite{omel2012nonuniversal,omel2013bifurcations}.
The phase lag induces a further drastic change that one bifurcation diagram includes both a continuous and a subsequent discontinuous transitions, where the latter emerges not from the incoherent state but from the partially synchronized state \cite{omel2012nonuniversal}.

Coming back to the original Kuramoto model, which is the target of this study,
we focus on the natural frequency distributions,
which have been assumed to be symmetric or unimodal so far.
By breaking the symmetry of distributions, the synchronization transition can change from continuous to discontinuous ones \cite{basnarkov2007phase,basnarkov2008kuramoto}.
However, asymmetric unimodal distributions have not generated a bifurcation diagram including a continuous and a discontinuous transitions,
which is reported in the Sakaguchi-Kuramoto model.

As discussed above, there are two ways to introduce asymmetry
in the Kuramoto model, but asymmetric distributions
have not been widely investigated
except for the limited cases
\cite{basnarkov2007phase,basnarkov2008kuramoto}.
Nevertheless, the asymmetric case should be considered,
since the symmetry of natural frequency distributions is not trivial
and slight asymmetry may induce qualitative changes in bifurcation diagrams.
Therefore, the first question considered in this article is:
Can we reproduce the nonstandard bifurcation diagram
which includes the two transitions and
is observed in the Sakaguchi-Kuramoto model
by introducing asymmetry of the natural frequency distribution
in the Kuramoto model?

Asymmetric unimodal distributions could not realize it yet.
Our idea is, therefore, to break the other assumption of unimodality,
and to consider asymmetric bimodal distributions.
Before going to the second question, we review bifurcation diagrams in the symmetric bimodal case.

As well as the asymmetric unimodal distributions, symmetric bimodal ones give discontinuous synchronization transitions \cite{bonilla1992nonlinear,crawford1994amplitude,martens2009exact,pazo2009existence,pietras2016coupled} but the prior continuous transition does not appear.
Apart from the stationary states and
extending objects to nonstationary states,
we have the collective oscillation with noise \cite{bonilla1992nonlinear,crawford1994amplitude} and later without noise \cite{martens2009exact,pazo2009existence,pietras2016coupled}.
This oscillatory state directly bifurcates
from the incoherent state as the partially synchronized state going through the continuous synchronization transition.
Thus, the second question arises: Is it possible for the oscillatory state to bifurcate from the partially coherent state as the subsequent discontinuous transition observed in the Sakaguchi-Kuramoto model?

The purposes of this article are to answer the above two questions
by introducing a family of smooth asymmetric bimodal distributions,
and to demonstrate that the asymmetry with the bimodality
in distributions plays a similar role to the asymmetry
in the coupling function.

A simple way to draw bifurcation diagrams is to perform $N$-body simulations, but the finite-size fluctuation is an obstacle.
The finite-size fluctuation is eliminated by taking the limit $N\to\infty$, and the Kuramoto model is described by the equation of continuity, but the infinite dimensionality is another difficulty.
To overcome this difficulty, we introduce the Ott-Antonsen ansatz \cite{ott2008low,ott2009long,ott2011comment}, which reduces the infinite-dimensional equation of continuity to a finite-dimensional dynamical system and succeeds in various oscillator systems \cite{martens2009exact,ott2008low,lee2009large,skardal2011cluster,tanaka2014solvable,pazo2014low,terada2014dynamics}.
The theoretical results obtained by the ansatz are verified by comparing with direct $N$-body simulations.

This article is organized as follows.
We introduce the Kuramoto model of coupled phase-oscillators in Sec. \ref{sec: model}.
The reduced system is derived by the Ott-Antonsen ansatz in Sec. \ref{sec:redction}.
The theoretical predictions in the reduced system are compared with the $N$-body simulations in Sec. \ref{sec:simulation}.
We summarize this article in the final section \ref{sec:conclusion}.

\section{The Kuramoto model}\label{sec: model}

The Kuramoto model is described by the equations
\begin{equation}
   \label{eq: kuramoto_model}
   \frac{\mathrm{d}\theta_i}{\mathrm{d}t}
   = \omega_i + \frac{K}{N}\sum_{j=1}^{N}\sin\left(\theta_j-\theta_i\right),
   \qquad (i=1,\cdots,N)
\end{equation}
where $\theta_i\in[0,2\pi)$ represents the phase of the $i$th oscillator and real $\omega_i$ the natural frequency.
The natural frequencies are obeyed by the probability density function $g\left(\omega\right)$, where $g\left(\omega\right)\mathrm{d}\omega$ is a fraction of oscillators having natural frequency between $\omega$ and $\omega+\mathrm{d}\omega$.
The coupling function is chosen as to be odd and $2\pi$-periodic,
and sine is one of the simplest choices.
We assumed that the interaction is all-to-all, and that
the coupling strength $K>0$ does not depend on coupling pairs.
The order parameter 
\begin{equation}
z_{N} = r_{N}e^{i\psi_{N}} = \frac{1}{N}\sum_{j=1}^{N}e^{i\theta_j}\label{eq: comp_order}
\end{equation}
measures the extent of synchronization by $r_N\ge0$ and the collective phase by $\psi_N$ for the $N$ oscillators.

In the limit of $N\to\infty$, we consider the probability density function $f\left(\theta,\omega,t\right)$, where $f\left(\theta,\omega,t\right)\mathrm{d}\theta\mathrm{d}\omega$ denotes the fraction of oscillators with phase between $\theta$ and $\theta+\mathrm{d}\theta$ and natural frequency between $\omega$ and $\omega+\mathrm{d}\omega$ at time $t$ under the normalization condition
\begin{equation}
\int^{\infty}_{-\infty}\mathrm{d}\omega\int^{2\pi}_{0}\mathrm{d}\theta f\left(\theta,\omega,t\right) = 1
\end{equation}
which induces
\begin{equation}
\int^{2\pi}_{0}\mathrm{d}\theta f\left(\theta,\omega,t\right) = g\left(\omega\right).
\end{equation}
Since the number of the oscillators conserves, the probability density function satisfies the equation of continuity:
\begin{equation}
\frac{\partial f}{\partial t}+\frac{\partial}{\partial\theta}\left(fv\right) = 0.\label{eq: con_eq}
\end{equation}
By defining the velocity field $v$ as
\begin{equation}
v\left(\theta,\omega,t\right) = \omega+K\int^{\infty}_{-\infty}\mathrm{d}\omega\int^{2\pi}_{0}\mathrm{d}\theta'\,f\left(\theta',\omega,t\right)\sin\left(\theta'-\theta\right), \label{eq: original_velocity}
\end{equation}
the equation of continuity corresponds to the evolution equations of the phase-oscillators (\ref{eq: kuramoto_model}) \cite{lancellotti2005vlasov}.

In the continuous limit, the order parameter is
\begin{equation}
  z = re^{i\psi} = \int^{\infty}_{-\infty}\mathrm{d}\omega\int^{2\pi}_{0}\mathrm{d}\theta'\,f\left(\theta',\omega,t\right)e^{i\theta}.\label{eq: order_param_inf}
\end{equation}
For a symmetric and unimodal $g\left(\omega\right)$,
after some calculations, Kuramoto derived the self-consistent equation
for $r$ which must be satisfied in stationary states \cite{kuramoto1984chemical}:
\begin{equation}
  \label{eq: self_consistent}
  %r = Kr\int^{\frac{\pi}{2}}_{-\frac{\pi}{2}}\mathrm{d}\theta\,g\left(Kr\sin\theta\right)\cos^2\theta.
  r = Kr\int^{\pi/2}_{-\pi/2}
  \mathrm{d}\theta\,g\left(Kr\sin\theta\right)\cos^2\theta.
\end{equation}
The self-consistent equation (\ref{eq: self_consistent})
has the unique trivial solution $r=0$ for $K\le K_c$,
but has one more positive solution for $K>K_c$.
The system exhibits a continuous spontaneous synchronization above $K_c$
where $K_c=2/\left[\pi g(0)\right]$
\cite{kuramoto1975self,strogatz2000kuramoto,gupta2014kuramoto,sasa2015collective}.
However, the breaking of symmetry or unimodality may change the continuity of transitions.
Indeed, asymmetric unimodal \cite{basnarkov2007phase,basnarkov2008kuramoto} and symmetric bimodal \cite{martens2009exact,pazo2009existence} cases
produce discontinuous transitions.

\begin{figure}[ht]
  \begin{center}
  \includegraphics{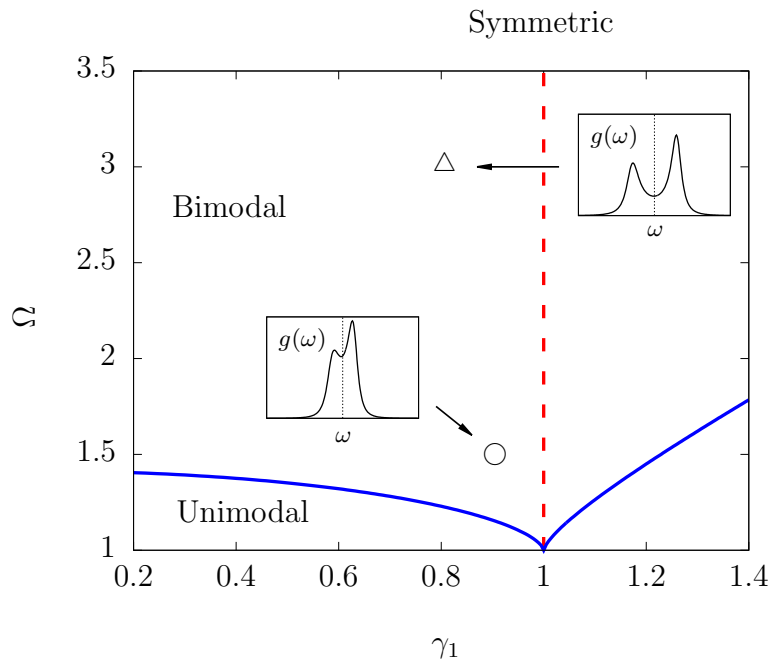}
    \caption{Classification of the family of distributions
      (\ref{eq: nat_fre_dist}) with $\gamma_2=1$.
      The red dashed line denotes the symmetric distributions.
      The blue solid lines are the borders
      between the unimodality and the bimodality.
      The circle and the triangle represent the parameter sets
      studied in Sec. \ref{sec:simulation}.
      (insets) Forms of the two distributions,
      where the dashed vertical lines denote $\omega=0$.}
\label{fig distribution_param}
\end{center}
\end{figure}

In this paper, to investigate the asymmetric bimodal case, we consider the following family of smooth natural frequency distributions:
\begin{equation}
  g\left(\omega\right)
  = \frac{c}{\left[\left(\omega-\Omega\right)^2+\gamma_1^2\right]\left[\left(\omega+\Omega\right)^2+\gamma_2^2\right]},\,\,\,\,\,(\Omega,\gamma_1,\gamma_2>0)\label{eq: nat_fre_dist}
\end{equation}
where
\begin{equation}
c = \frac{\gamma_1\gamma_2\left[\left(\gamma_1+\gamma_2\right)^2+4\Omega^2\right]}{\pi\left(\gamma_1+\gamma_2\right)}
\end{equation}
is the normalization constant.
This family, whose form is the product of two Lorentzian distributions,
can be  bimodal and/or asymmetric depending on parameters $\Omega$, $\gamma_1$,
and $\gamma_2$.
We note that the Kuramoto model (\ref{eq: kuramoto_model}) is invariant
under appropriate scalings of $t, \omega_i,K,\Omega,\gamma_1$ and $\gamma_2$.
Thus, we impose the arbitrarity of $\gamma_2$ to the other parameters
and set $\gamma_2=1$ without loss of generality.
Figure \ref{fig distribution_param} represents the diagram
of the distribution forms.
We explore only the left-half plane from the symmetry line
without loss of generality, since bifurcations in the other side
are replicated by putting $\theta\to-\theta$.

\section{The Ott-Antonsen reduction}\label{sec:redction}

The Ott-Antonsen ansatz \cite{ott2008low,ott2009long,ott2011comment} is a powerful tool to reduce the infinite-dimensional equations of continuity to finite-dimensional ordinary differential equations.
In practice, many works succeed in reducing the dynamics of the coupled phase-oscillators
\cite{martens2009exact,ott2008low,lee2009large,skardal2011cluster,tanaka2014solvable,pazo2014low,terada2014dynamics}.
In this section, we employ the ansatz and derive the reduced equation to obtain precise bifurcation diagrams.

The Ott-Antonsen ansatz \cite{ott2008low} introduces the Fourier expansion of $f\left(\theta,\omega,t\right)$ as
\begin{equation}
f\left(\theta,\omega,t\right) = \frac{g\left(\omega\right)}{2\pi}\left[1+\sum_{n=1}^{\infty}\left(a\left(\omega,t\right)^ne^{in\theta}+\bar{a}\left(\omega,t\right)^ne^{-in\theta}\right)\right],\label{eq: pdf_ansatz}
\end{equation}
where we assume $|a\left(\omega,t\right)|<1$ to ensure convergence of the sum.
From the assumption (\ref{eq: pdf_ansatz})
and the continuous equation (\ref{eq: con_eq}),
the Kuramoto model with the distribution (\ref{eq: nat_fre_dist})
is reduced to the system of the two complex variables $A$ and $B$
(see the Appendix A for the derivation):
\begin{eqnarray}
  \frac{\mathrm{d}A}{\mathrm{d}t} &= \left(i\Omega-\gamma_1\right)A-\frac{K}{2}\left[\left(\bar{k}_1\bar{A}+\bar{k}_2\bar{B}\right)A^2-\left(k_1A+k_2B\right)\right],\label{eq: reduced1}\\
  \frac{\mathrm{d}B}{\mathrm{d}t} &= -\left(i\Omega+\gamma_2\right)B-\frac{K}{2}\left[\left(\bar{k}_1\bar{A}+\bar{k}_2\bar{B}\right)B^2-\left(k_1A+k_2B\right)\right],\label{eq: reduced2}
\end{eqnarray}
where we set the constant parameters $k_1$ and $k_2$ as 
\begin{equation}
k_1 = \frac{\gamma_2\left[2\Omega-i\left(\gamma_1+\gamma_2\right)\right]}{\left(\gamma_1+\gamma_2\right)\left[2\Omega+i\left(\gamma_1-\gamma_2\right)\right]}, \,\,\,\,\,
k_2 = \frac{\gamma_1\left[2\Omega+i\left(\gamma_1+\gamma_2\right)\right]}{\left(\gamma_1+\gamma_2\right)\left[2\Omega+i\left(\gamma_1-\gamma_2\right)\right]}.
\end{equation}
We have fixed as $\gamma_{2}=1$, but the symbol $\gamma_{2}$ remains to clarify how it appears in the above equations.
We note that the relation $k_2=\bar{k}_1$ holds if the distribution is symmetric, namely $\gamma_1=\gamma_2$.
The complex order parameter $z$ is represented by $A$ and $B$ as
\begin{equation}
z = k_1A+k_2B,
\end{equation}
and we concentrate on the amplitude $r=|z|$, which expresses the extent of the synchronization.

We construct bifurcation diagrams by searching steady states
in which the amplitude $r(t)$ is constant or oscillatory.
The former and the latter states in the reduced system
correspond to the stationary states and the oscillatory states, respectively
in the original system.
The formers are further classified into the trivial fixed-point solution
of $A=B=0$ corresponding to the incoherent state with $r=0$,
and nontrivial solutions with constant $r>0$
corresponding to the partially synchronized states.
Stability of the trivial solution is analyzed by considering
the $4\times 4$ Jacobian matrix derived from
Eqs. (\ref{eq: reduced1}) and (\ref{eq: reduced2}).
For detecting the nontrivial solutions,
we perform one more reduction since $r$ depends on $r_A,r_B$ and $\phi$ only,
where we introduced the polar forms $A=r_Ae^{-i\phi_A}, B=r_Be^{-i\phi_B}$
and the phase difference variable $\phi=\phi_A-\phi_B$.
The three variables are obeyed by the equations
\begin{eqnarray}
\frac{\mathrm{d}r_A}{\mathrm{d}t} &=& -\gamma_1r_A+\frac{K}{2}\left(1-r_A^2\right)\left(\mathrm{Re}k_1r_A+\mathrm{Re}k_2r_B\cos\phi+\mathrm{Im}k_2r_B\sin\phi\right),\label{eq: redued_polar1}\\
\frac{\mathrm{d}r_B}{\mathrm{d}t} &=& -\gamma_2r_B+\frac{K}{2}\left(1-r_B^2\right)\left(\mathrm{Re}k_1r_A\cos\phi-\mathrm{Im}k_1r_A\sin\phi+\mathrm{Re}k_2r_B\right),\label{eq: redued_polar2}\\
\frac{\mathrm{d}\phi}{\mathrm{d}t} &=& 2\Omega+\frac{K}{2}\Biggl[\left(1+r_A^2\right)\left(\mathrm{Im}k_1+\mathrm{Im}k_2\frac{r_B}{r_A}\cos\phi-\mathrm{Re}k_2\frac{r_B}{r_A}\sin\phi\right)\nonumber\\
&&-\left(1+r_B^2\right)\left(\mathrm{Im}k_2+\mathrm{Im}k_1\frac{r_A}{r_B}\cos\phi+\mathrm{Re}k_1\frac{r_A}{r_B}\sin\phi\right)\Biggr],\label{eq: redued_polar3}
\end{eqnarray}
where $\mathrm{Re}\,k_i$ and $\mathrm{Im}\,k_i$ represent
the real and the imaginary parts of $k_i$ $(i=1,2)$, respectively.
The nontrivial solutions are found as the fixed points
of the $3$-dimensional system,
and their stabilities are sufficiently captured
by the associated $3\times 3$ Jacobian matrix.
Based on this discussion, we search the oscillatory states
by performing numerical simulations of the $3$-dimensional system.

\section{Bifurcation diagrams}\label{sec:simulation}

In this section we obtain the two new types of bifurcation diagrams by computing fixed points of the reduced equations (\ref{eq: redued_polar1})-(\ref{eq: redued_polar3}) theoretically, and oscillating solutions numerically.
The bifurcation diagrams are validated by comparing with direct $N$-body simulations of Eq. (\ref{eq: kuramoto_model}).
All the simulations of the reduced system and the $N$-body system are performed by the fourth-order Runge-Kutta method with the time step $dt=0.01$.
The synchronization and the collective oscillation are investigated by calculating the averages and the standard deviations of $r$ in the time interval $t\in[4500,5000]$, in which we confirmed that the system reaches a steady state.
In section \ref{sec:fixed_points}, we show the first bifurcation diagram consisting of stationary states corresponding to the fixed points of the reduced equations (\ref{eq: redued_polar1})-(\ref{eq: redued_polar3}).
In section \ref{sec:standing_waves}, the second bifurcation diagram is reported which includes the oscillatory states corresponding to the limit-cycle solutions of the reduced system.

\subsection{Bifurcations of stationary states}\label{sec:fixed_points}

\begin{figure}[h]
\begin{center}
  \includegraphics{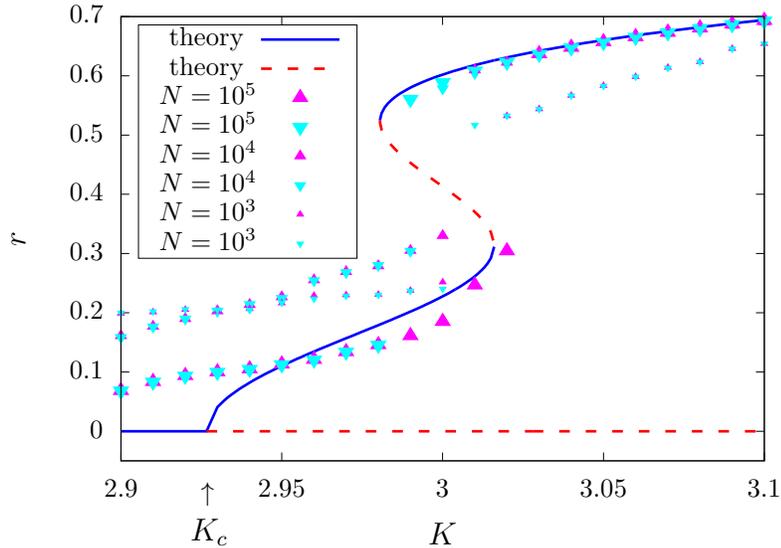}
\caption{
Bifurcations of stationary states in the Kuramoto model for $\left(\Omega,\gamma_1\right)=\left(1.5,0.9\right)$.
The lines are theoretically obtained fixed-point solutions to the reduced equations (\ref{eq: redued_polar1})-(\ref{eq: redued_polar3}).
The blue solid and the red dashed lines represent the stable and the unstable branches, respectively.
The points are the time averages in the $N$-body simulations.
The triangles and the inverse triangles indicate the forward and the backward processes, respectively, where the forward process increases $K$ with $\mathrm{d}K=0.01$ whereas the backward process decreases $K$ after the forward simulation.
The sizes of the points indicate the system sizes.
}
\label{fig analytical_bifurcation}
\end{center}
\end{figure}

We here concentrate on exploring nontrivial fixed-point solutions by setting $\mathrm{d}r_A/\mathrm{d}t=\mathrm{d}r_B/\mathrm{d}t=\mathrm{d}\phi/\mathrm{d}t=0$ in Eqs. (\ref{eq: redued_polar1})-(\ref{eq: redued_polar3}).
Analytical solutions for the parameter set $\left(\Omega,\gamma_1\right)=\left(1.5,0.9\right)$, which makes $g(\omega)$ asymmetric and bimodal, are plotted as lines in Fig. \ref{fig analytical_bifurcation} with the trivial solution $r=0$.
We marked $K_c$ derived theoretically from the equation of continuity (see the Appendix B)
at which the incoherent state changes the stability.
The derived $K_c$ is in good agreement with the analysis
of the reduced system.
At $K_c$, the stable incoherent state continuously bifurcates
into the unstable incoherent state and the partially synchronized state
as the symmetric unimodal case.
Further increasing $K$, we find the discontinuous transition
and coexistence of the two partially synchronized stable branches.
The subsequent discontinuous transition is observed
in the Sakaguchi-Kuramoto model \cite{omel2012nonuniversal},
but according to our best knowledge, this type of bifurcation diagram
has not been discovered in the Kuramoto model.

To verify the bifurcation diagram obtained under the Ott-Antonsen ansatz, we numerically simulate the $N$-body system.
Here, we set $N=10^3, 10^4, 10^5$ and integrate Eq. (\ref{eq: kuramoto_model}) with natural frequencies randomly generated from the distribution Eq. (\ref{eq: nat_fre_dist}).
To see the hysteresis in the discontinuous transition the simulations are performed  with firstly increasing and subsequently decreasing the coupling strength $K$, where we call the former process with the increase of $K$ the forward process and the latter the backward process, respectively.
The forward simulation denoted as the triangles in Fig. \ref{fig analytical_bifurcation} increases $K$ with the step $\mathrm{d}K=0.01$ from $K=2.9$ to $K=3.1$ and the initial state is set to the previous final state at each $K$ except for $K=2.9$ in which the initial conditions of the phase variables are chosen randomly from the uniform distribution on the circle.
The backward process is represented as the inverse triangles in Fig. \ref{fig analytical_bifurcation}, which is realized by decreasing in the same interval $[2.9,3.1]$ of $K$ and the step $\mathrm{d}K=0.01$.
All the numerically computed points, and the discontinuous transition in particular, are qualitatively in good agreements with the Ott-Antonsen analysis.
Consequently, the theoretically obtained new bifurcation diagram has been verified.

\subsection{Bifurcations with oscillatory states}\label{sec:standing_waves}

Previous works on the Kuramoto model with symmetric bimodal distributions \cite{martens2009exact,pazo2009existence} reported that the reduced system has not only the fixed-point solutions but also limit-cycle solutions, which correspond to the collective oscillations.
We search oscillatory states in our asymmetric bimodal case by performing numerical simulations of the reduced system.
The main strategy is to increase the parameter $\Omega$ representing the distance between the two peaks of the distribution $g(\omega)$ from the value examined in Sec. \ref{sec:fixed_points}.

\begin{figure}[h]
\begin{center}
  \includegraphics{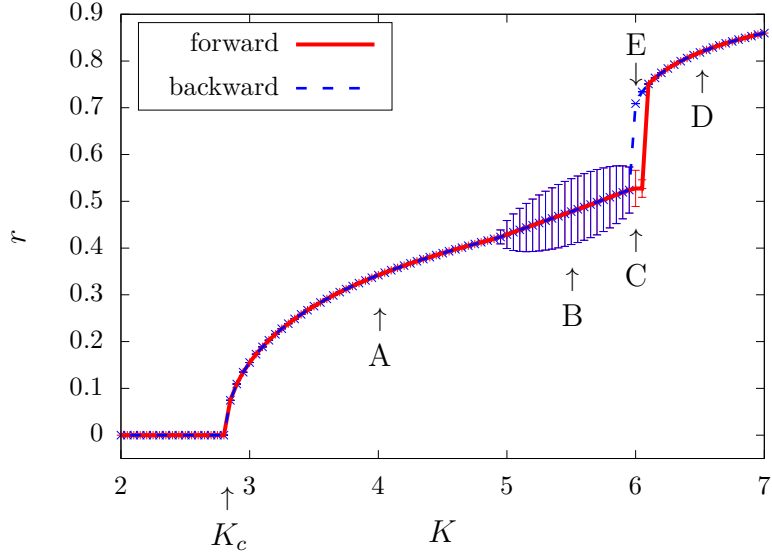}
\caption{
Bifurcation diagram for $\left(\Omega,\gamma_1\right)=\left(3.0,0.8\right)$ obtained by simulating the reduced $3$-dimensional system.
The averages of the order parameter are plotted by points, which are connected with lines to emphasize the continuous and the discontinuous transitions at $K=K_{c}$ and around $K=6$, respectively.
The standard deviations are represented by vertical bars associated with the points, where large deviations indicate oscillation of the order parameter.
Red (blue) points and lines are for the forward (backward) process, where the forward process increases $K$ with $\mathrm{d}K=0.05$ whereas the backward process decreases $K$ after the forward simulation.
For the points A, B, C, D and E, temporal evolutions of the order parameter are plotted in Fig. \ref{fig non-stationary2}, and synchronizations on the $(\theta,\omega)$ plane in Fig. \ref{fig snapshots}.
}
\label{fig non-stationary}
\end{center}
\end{figure}

For the parameter set $(\Omega,\gamma_{1})=(3.0, 0.8)$, the averages and the standard deviations of $r(t)$ are plotted in Fig.\ref{fig non-stationary} for the steady states realized after transient durations.
Large standard deviations indicate the oscillation of $r$, and the oscillatory states are observed in our asymmetric bimodal case around the interval $[5,6]$ of $K$.

We give two remarks on the bifurcation diagram.
First, the discontinuous transition emerges from the oscillatory states, and the hysteresis is confirmed by performing the forward and the backward processes as done in the previous section \ref{sec:fixed_points} but for the reduced system with the step $\mathrm{d}K=0.05$, where we set $r_A,r_B\sim0$ and  $\phi$ to the arbitrary constant as the initial state of the forward process and give the perturbation to $r_A,r_B$ at each $K$ to see the stability.
Accordingly, the coexistence of an oscillatory and a stationary states is observed as in the symmetric bimodal case \cite{martens2009exact,pazo2009existence,pietras2016coupled}.
Second, the oscillatory state appear after the continuous transition, and arise from the partially synchronized states around $K=5$.
This emergence manner of the collective oscillation shows a sharp contrast with the symmetric bimodal case, in which it arises from the incoherent state.

\begin{figure}[h]
\begin{center}
  \includegraphics{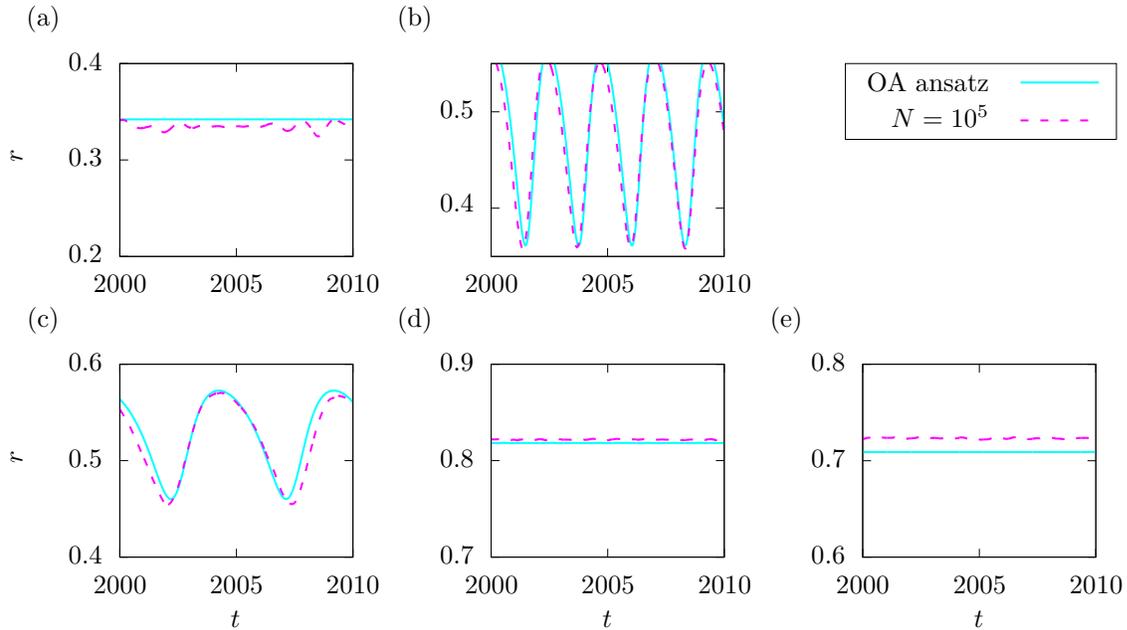}
\caption{
Comparison of time evolutions of the order parameters between the reduced system (solid light blue lines) and $N$-body system (dashed magenta lines) for $\left(\Omega,\gamma_1\right)=\left(3.0,0.8\right)$.
The examined points are marked in Fig. \ref{fig non-stationary}: (a) A, (b) B, (c) C, (d) D from the forward process and (e) E from the backward process.
}
\label{fig non-stationary2}
\end{center}
\end{figure}

We examine the bifurcation diagram obtained in the reduced system by performing $N$-body simulations at the points A ($K=4.0$), B ($K=5.5$), C ($K=6.0$), D ($K=6.5$) in the forward process and E ($K=6.0$) in the backward process marked in Fig. \ref{fig non-stationary}.
At each point, temporal evolutions of $r(t)$ are compared between the reduced system and the $N$-body system in Fig. \ref{fig non-stationary2}.
For the comparisons, we shifted the $N$-body lines properly, since the initial phase of the collective oscillation do not matter.
The two lines show good agreements, even under the finite-size fluctuation.

\begin{figure}[h]
\begin{center}
  \includegraphics{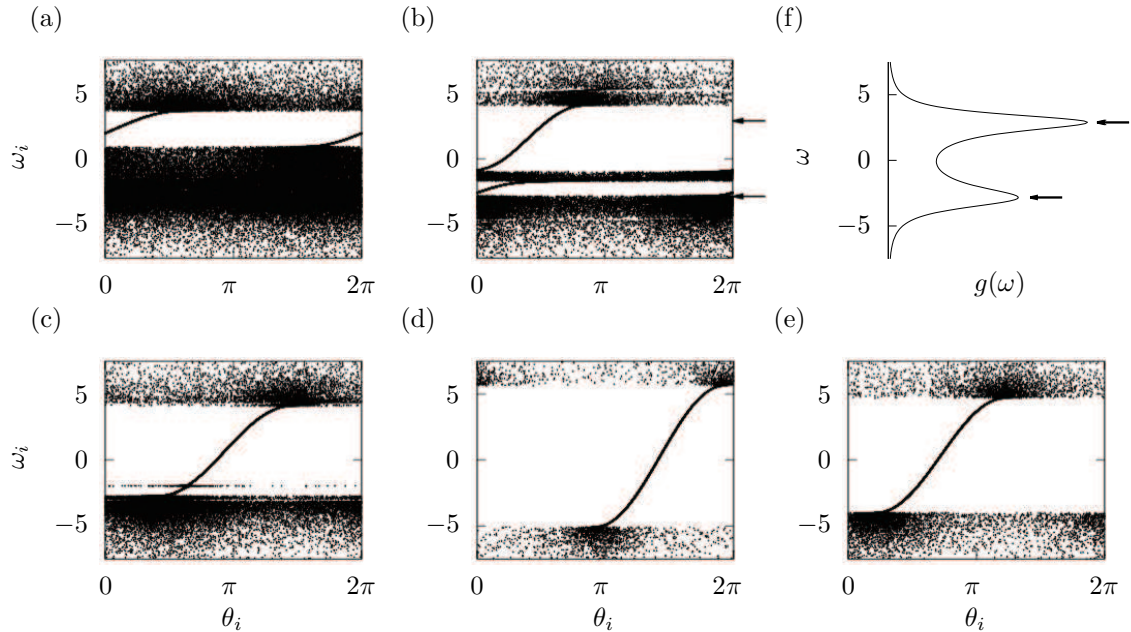}
\caption{
Snapshots of $\{(\theta_i,\omega_i)\}$ in the $N$-body simulations for $\left(\Omega,\gamma_1\right)=\left(3.0,0.8\right)$ with $N=10^5$.
The examined points are marked in Fig. \ref{fig non-stationary}: (a) A, (b) B, (c) C, (d) D from the forward process and (e) E from the backward process.
The natural frequency distribution is displayed in the panel (f) for comparing with the panel (b) in particular.
The two peaks of the distribution are pointed to by arrows in the panels (b) and (f).
}
\label{fig snapshots}
\end{center}
\end{figure}

To look into the oscillatory states observed here, in Fig. \ref{fig snapshots}
we represent the snapshots of $\{(\theta_i,\omega_i)\}$ in the steady states of the $N$-body simulations at the five points of A, B, C, D and E.
The population of the oscillators forms only one synchronized cluster in Figs. \ref{fig snapshots} (a), (d) and (e), but the two clusters exist in Figs. \ref{fig snapshots} (b) and (c).
We observe that existence of the second smaller cluster induces the oscillating states, and coalescing of the two cluster triggers the discontinuous transition. 
If we had a symmetric bimodal natural frequency distribution, the two groups of the synchronized oscillators would emerge simultaneously as seen in \cite{martens2009exact,pazo2009existence,pietras2016coupled}, and the oscillatory states would emerge at $K_c$ from the incoherent state.
However, the asymmetry permits the two groups to synchronize separately, and provides the continuous synchronization transition at $K_c$.
Another interesting observation is that the natural frequencies in the second cluster are not located at the second peak of the natural frequency distribution as shown in Fig. \ref{fig snapshots} (f).

\section{Conclusion and discussion}\label{sec:conclusion}

We studied bifurcation diagrams in the Kuramoto model having asymmetric bimodal distributions of natural frequencies.
With the aid of the Ott-Antonsen ansatz \cite{ott2008low}, we theoretically found the two new bifurcation diagrams:
One includes continuous and subsequent discontinuous transitions and the other shows the oscillatory state bifurcating from the partially coherent states.
These theoretical results were validated by comparing with direct $N$-body simulations.

The first diagram we found has been reported in the Sakaguchi-Kuramoto model with the symmetric unimodal distribution.
However, according to our best knowledge, such a bifurcation diagram has not been observed in the Kuramoto model, and this is the first observation by introducing asymmetric bimodal distributions.

The second diagram has the oscillatory states, in which the magnitude of the order parameter oscillates.
With the symmetric bimodal distributions the oscillatory states have been reported to bifurcate from the incoherent state \cite{martens2009exact,pazo2009existence,pietras2016coupled}.
However, the asymmetry lets the oscillatory state emerge from partially synchronized state in the bifurcation diagram.
This emergence happens because the synchrony begins around the higher peak, which forms the first cluster, and the second synchrony subsequently occurs, which forms the relatively traveling second cluster as discussed in the section \ref{sec:standing_waves}.
It is worth noting that the second synchrony is not located
at the center of the lower peak.

From the perspective obtained by the second diagram, we reconsider the first diagram.
Contrastingly with the second bifurcation diagram, we numerically confirmed by the $N$-body simulations that the first diagram has no second cluster.
We suppose that, as the separation of the two peaks become smaller, the oscillatory states tend to disappear and the bifurcation diagram approaches from the second bifurcation diagram to the first.

We give the two remarks to emphasize the worth of this article.
First, Kuramoto sketched the successive synchronies observed in the second bifurcation diagram for the asymmetric bimodal case in \cite{kuramoto1984chemical}, where the first and the second synchronies were expected to appear around the higher and the lower peaks of $g(\omega)$, respectively, and to be isolated.
Beyond the sketch, this article has revealed that the second cluster may not be located around the center of the lower peak as shown in Fig. \ref{fig snapshots}, and may not appear as in the first diagram.
Second, we demonstrated that the asymmetry of natural frequency distributions discloses the bifurcation diagrams hidden in the symmetric case, as the Sakaguchi-Kuramoto model can do by introducing asymmetry in the coupling function.
Indeed, keeping the peak-separation parameter $\Omega$ and setting the distribution to be symmetric, we could not find the reported new bifurcation diagrams.
We, therefore, conclude that the asymmetry in the natural frequency distributions plays a similar role to the asymmetry in the coupling function, introduced in the Sakaguchi-Kuramoto model.

We further discuss on the asymmetry by using analogies between the Kuramoto model and Hamiltonian systems.
We start from referring to  two analogies.
The Kuramoto model shares macroscopic aspects with a Hamiltonian system in the critical exponents concerning with the applied external force (see \cite{hong2007entrainment,nishikawa2014finite,hong2015finite,daido2015susceptibility} for the Kuramoto model and \cite{patelli2012linear,ogawa2014non,ogawa2014nonlinear,ogawa2015landau,yamaguchi2016strange} for the Hamiltonian system).
Another analogy is used in the analysis on the Landau damping \cite{landau1946vibrations}, which is originally studied in plasma physics, and is applied to the Kuramoto model \cite{strogatz1992coupled}.
Taking account of the above analogies, the analyticity of natural frequency distributions is expected to change macroscopic dynamics in the Kuramoto model, since non-analytic momentum distributions in a Hamiltonian system, which corresponds to the natural frequency distributions in the Kuramoto model, make the exponential Landau damping algebraic \cite{weitzner1963plasma}.
This perspective tells us that the study on the asymmetric unimodal case has not completed, because the previously studied distributions are not differentiable \cite{basnarkov2007phase,basnarkov2008kuramoto}. 
Our preliminary researches did not find nonstandard bifurcation diagrams in the analytic asymmetric unimodal case, but further investigations are necessary.
It is still an open question what is the whole role of asymmetry.
For instance, an asymmetric unimodal momentum distribution makes the susceptibility tensor  non-diagonable even in a simple Hamiltonian system \cite{yamaguchi2015nondiagonalizable}.
Thus, it might be interesting to study the response to an external force in the Kuramoto model with asymmetric natural frequency distributions.

\section*{Acknowledgement}

YT, KI and TA are supported by MEXT KAKENHI Grant Numbers 15H05877 and 26120006, and by JSPS KAKENHI Grant Numbers 16KT0019 and 15587273.
YYY is supported by JSPS KAKENHI Grant Numbers 23560069 and 16K05472.

\section*{Appendix A. Derivation of the reduced equations}

We derive the reduced equations (\ref{eq: reduced1}) and (\ref{eq: reduced2}) using the Ott-Antonsen ansatz from the original equation (\ref{eq: kuramoto_model}) and the natural frequency distribution (\ref{eq: nat_fre_dist}).

We consider the Kuramoto model in the continuous limit $N\to\infty$, and from Eq. (\ref{eq: kuramoto_model}) the velocity of the phase can be written as
\begin{equation}
v\left(\theta,\omega,t\right) = \omega+\frac{K}{2i}\left(ze^{-i\theta}-\bar{z}e^{i\theta}\right),\label{eq: velocity_function}
\end{equation}
where  we denote the complex order parameter (\ref{eq: order_param_inf}) as $z(t)=re^{i\psi}$.
The Ott-Antonsen ansatz \cite{ott2008low} assumes
that the probability distribution function $f\left(\theta,\omega,t\right)$
is expanded into the Fourier series as Eq.  (\ref{eq: pdf_ansatz}).
Substituting the assumed $f\left(\theta,\omega,t\right)$ (\ref{eq: pdf_ansatz}) and the velocity (\ref{eq: velocity_function}) into the continuous equation (\ref{eq: con_eq}), we obtain the equation for $a\left(\omega,t\right)$ and $z\left(t\right)$ as
\begin{equation}
\frac{\partial a}{\partial t}+i\omega a+\frac{K}{2}\left(a^2z-\bar{z}\right) = 0.\label{eq: ev_a_z}
\end{equation}
From Eqs. (\ref{eq: order_param_inf}) and (\ref{eq: pdf_ansatz}), we also find
\begin{equation}
z(t) = \int^{\infty}_{-\infty}\mathrm{d}\omega\,g\left(\omega\right)\bar{a}\left(\omega,t\right).\label{eq: order_g_a}
\end{equation}
From the complex conjugate of Eq. (\ref{eq: ev_a_z}), $\partial\bar{a}/\partial t\sim -\left(\mathrm{Im}\,\omega\right)\bar{a}$ as $\mathrm{Im}\,\omega\to\infty$.
We then find that $\bar{a}\left(\omega,t\right)\to0$ as $\mathrm{Im}\,\omega\to\infty$. % from the assumption $|a\left(\omega,t\right)|<1$.
The vanishing $\bar{a}\left(\omega,t\right)$ permits us to add the upper semicircular contour to the integration of Eq. (\ref{eq: order_g_a}) in the complex $\omega$ plane for using the residue theorem.
Thus, assuming analyticity of $\bar{a}\left(\omega,t\right)$ in $\mathrm{Im}\,\omega>0$ \cite{ott2008low}, we have
\begin{equation}
z(t) = k_1\bar{a}\left(\Omega+i\gamma_1,t\right)+k_2\bar{a}\left(-\Omega+i\gamma_2,t\right),
\end{equation}
because the distribution (\ref{eq: nat_fre_dist}) has two poles
at $\Omega+i\gamma_1$ and $-\Omega+i\gamma_2$ in the upper half plane.
Temporal evolutions of the complex variables
\begin{equation}
A(t) = \bar{a}\left(\Omega+i\gamma_1,t\right),\,\,\, B(t) = \bar{a}\left(-\Omega+i\gamma_2,t\right),
\end{equation}
are obtained from the complex conjugate of Eq. (\ref{eq: ev_a_z}) by setting $\omega$ as $\omega=\Omega+i\gamma_1$ and $\omega=-\Omega+i\gamma_2$, respectively, and are expressed by Eqs. (\ref{eq: reduced1}) and (\ref{eq: reduced2}).

\section*{Appendix B. Calculation of $K_c$ from the linear stability analysis in the continuum equation}

We here derive the equation determining the critical coupling strength $K_c$, at which the incoherent solution to the equation of continuity (\ref{eq: con_eq})
\begin{equation}
f_0\left(\theta,\omega\right) = \frac{g(\omega)}{2\pi} \label{eq: incoherent_solution}
\end{equation}
is destabilized, where $g(\omega)$ is the natural frequency distribution.
The critical strength $K_c$ is obtained by the linear stability analysis.
We consider a perturbed solution $f(\theta,\omega,t)$ with 
a perturbation $\epsilon f_1\left(\theta,\omega,t\right)$ as
\begin{equation}
f \left(\theta,\omega,t\right)= f_0\left(\theta,\omega\right)+\epsilon f_1\left(\theta,\omega,t\right), \label{eq: f_0andf_1}
\end{equation}
where $\left|\epsilon\right|\ll1$, and investigate when the perturbation grows. 

We explicitly write the velocity functional $v[f]$ (\ref{eq: original_velocity}) as
\begin{eqnarray}
v\left[f\right] = \omega+V\left[f\right],\,\,\,\,\,V[f]\equiv K\int^{\infty}_{-\infty} d\omega\int^{2\pi}_{0} d\theta'f\left(\theta',\omega,t\right)\sin\left(\theta'-\theta\right).\label{eq: velocity_functional}
\end{eqnarray}
Substituting the expansions of $f$ (\ref{eq: f_0andf_1}) and $v$ (\ref{eq: velocity_functional}) into the equation of continuity,
and picking up the terms of $O(\epsilon)$,
%the equation in the first order of $\epsilon$,
we have
\begin{equation}
\frac{\partial}{\partial t}f_1+\frac{\partial}{\partial\theta}\left(v_0f_1+V_1f_0\right) = 0,\label{eq: con_eq_pert}
\end{equation}
where we put $v_0=v\left[f_0\right]$ and $V_1=V\left[f_1\right]$.
Expanding each function for the Fourier series as
\begin{equation}
f_1\left(\theta,\omega,t\right) = \sum_{n=-\infty}^{\infty}\tilde{f}_1\left(n,\omega,t\right)e^{in\theta},
\end{equation} 
we obtain the equation
\begin{equation}
\frac{\partial\tilde{f}_1}{\partial t}\left(n,\omega,t\right) = -in\left(\tilde{f}_1\left(n,\omega,t\right)\tilde{v}_0\left(0,\omega\right)+\tilde{f}_0\left(0,\omega\right)\tilde{V}_1\left(n,\omega,t\right)\right),
\end{equation}
where we utilized the fact $\tilde{v}_0\left(n,\omega\right)=\tilde{f}_0\left(n,\omega\right)=0$ for nonzero integer $n$.
Performing the Laplace transforms
\begin{equation}
\hat{f}_{\blue{1}}\left(n,\omega,s\right) = \int^{\infty}_{0}dt\,\tilde{f}_1\left(n,\omega,t\right)e^{-st},
\end{equation}
where Re$\,s>0$ to ensure the convergence, we have
\begin{equation}
\left(s+in\tilde{v}_0\left(0,\omega\right)\right)\hat{f}_1\left(n,\omega,s\right) = \tilde{f}_1\left(n,\omega,0\right)-in\tilde{f}_0\left(0,\omega\right)\hat{V}_1\left(n,\omega,s\right).\label{continuum_laplace}
\end{equation}
We consider the cases $n\neq\pm1$ and $n=\pm1$ separately.
For the case $n\neq\pm1$ it is easily shown that $\tilde{V}(n,\omega,t)=0$ and from Eq. (\ref{continuum_laplace}) we obtain
\begin{equation}
\hat{f}_{\blue{1}}\left(n,\omega,s\right) = \frac{\tilde{f}_1\left(n,\omega,0\right)}{s+in\omega}.\label{eq: hat_f}
\end{equation}
Since we see $\tilde{f}_1\left(n,\omega,t\right)=\tilde{f}_1\left(n,\omega,0\right)e^{-in\omega t}$ from the inverse Laplace transform of $\hat{f}_{\blue{1}}\left(n,\omega,s\right)$ with Eq. (\ref{eq: hat_f}), it is shown that the magnitude of $\tilde{f}_1\left(n,\omega,t\right)$ does not change with respect to time $t$ as
\begin{equation}
\left|\tilde{f}_1\left(n,\omega,t\right)\right| = \left|\tilde{f}_1\left(n,\omega,0\right)\right|.
\end{equation}
This computation implies that the modes for $n\neq\pm1$ do not contribute to the destabilization of the incoherent solution, and we then focus on the case $n=\pm1$.
In the light of the fact that $|\tilde{f}_1\left(1,\omega,t\right)| = |\tilde{f}_1\left(-1,\omega,t\right)|$, it is enough to consider only the case $n=1$.
For $n=1$ the equation (\ref{continuum_laplace}) becomes
\begin{equation}
\left(s+i\omega\right)\hat{f}_1\left(1,\omega,s\right) = \tilde{f}_1\left(1,\omega,0\right)+\pi K\tilde{f}_0\left(0,\omega\right)\int^{\infty}_{-\infty}\mathrm{d}\omega\,\hat{f}_1\left(1,\omega,s\right).\label{eq: continuum_laplace_n1}
\end{equation}
We define the function
\begin{equation}
\hat{\rho}_1\left(n,s\right) = \int^{\infty}_{-\infty}\mathrm{d}\omega\,\hat{f}_1\left(n,\omega,s\right),
\end{equation}
and integrate Eq. (\ref{eq: continuum_laplace_n1}) over $\omega$
from $-\infty$ to $\infty$.
Then, we formally obtain
\begin{equation}
  \hat{\rho}_1\left(1,s\right) = 
  \frac{1}{D_{K}(s)}
  %\left(1-\pi K\int^{\infty}_{-\infty}\mathrm{d}\omega\,\frac{\tilde{f}_0\left(0,\omega\right)}{s+i\omega}\right)^{-1}
  \int^{\infty}_{-\infty}\mathrm{d}\omega\,\frac{\tilde{f}_1\left(1,\omega,0\right)}{s+i\omega},
\end{equation}
where the function $D_{K}(s)$ is defined by
\begin{eqnarray}
  D_K\left(s\right)
  &=1+i\frac{K}{2}\int^{\infty}_{-\infty}\mathrm{d}\omega\,\frac{g\left(\omega\right)}{\omega-is}
  % &=1+i\frac{K}{2}\int^{\infty}_{-\infty}d\omega \frac{g\left(\omega\right)}{\omega-is},
  \label{eq: function_D}
\end{eqnarray}
in the domain Re$\,s>0$ by recalling the domain of the Laplace transform and $\tilde{f}_0\left(0,\omega\right)=g\left(\omega\right)/(2\pi)$.
A root of $D_K\left(s\right)$, denoted by $s_0$,
gives a pole of $\hat{\rho}_1\left(1,s\right)$ at $s=s_0$,
and this pole gives the estimation of $\tilde{\rho}_{1}(1,t)\sim e^{s_0t}$
in the linear regime by the inverse Laplace transform
of $\hat{\rho}_1\left(1,s\right)$.
Therefore, roots of $D_K\left(s\right)$ in the domain Re$\,s>0$ imply instability of the incoherent state $f_0(\theta,\omega)$ (\ref{eq: incoherent_solution}).
The strongest instability is given by the root of $D_K\left(s\right)$ whose real part is the maximum, and we denote the root by $s^{\ast}(K)$ depending on $K$.
The critical strength $K_{c}$ is determined by the condition $\lim_{K\to K_{c}}$Re$\,s^{\ast}(K)=0$.
A benefit of the Lorentzian type $g(\omega)$ (\ref{eq: nat_fre_dist}) is that we can conduct explicitly the integral in Eq. (\ref{eq: function_D}) with the aid of the residue theorem and obtain the critical coupling strength $K_{c}$ exactly.

\section*{References}

%\bibliography{iopart-num}
\bibliographystyle{unsrt}
\bibliography{asymmetry_ref}

\end{document}